\documentclass[]{article}
\usepackage[utf8]{inputenc} 
\usepackage{authblk}
\usepackage{geometry} 
\usepackage{graphicx} 
\graphicspath{{./pictures/}} 
\usepackage{gensymb} 
\usepackage{booktabs} 
\usepackage{array} 
\usepackage{paralist} 
\usepackage{verbatim} 
\usepackage{subfig} 
\usepackage{gensymb} 
\usepackage{booktabs} 
\usepackage{amsmath}
\usepackage{amssymb}
\usepackage{amsthm}
\usepackage{amsfonts}
\usepackage{braket}
\usepackage[euler]{textgreek}
\usepackage{xcolor}

\usepackage{fancyhdr} 
\usepackage{sectsty}
\allsectionsfont{\sffamily\mdseries\upshape} 

\usepackage[nottoc,notlof,notlot]{tocbibind} 
\usepackage[titles,subfigure]{tocloft} 

\title{Quantum Diamond Radio Frequency Signal Analyser based on Nitrogen-Vacancy centers}

\author[1]{Simone Magaletti}

\author[1]{Ludovic Mayer}

\author[2]{Jean-François Roch}

\author[1,*]{Thierry Debuisschert}

\affil[1]{{\small Thales Research and technology, 1 Avenue Augustin Fresnel, Palaiseau Cedex, 91767,France}}

\affil[2]{{\small Université Paris-Saclay, CNRS, ENS Paris-Saclay, CentraleSupelec, LuMIn,Gif-sur-Yvette, 91190, France}}

\affil[*]{{\small thierry.debuisschert@thalesgroup.com}}

\begin{document}
\date{}
\maketitle

\begin{abstract}
The fast development of radio-frequency (RF) technologies increases the need for compact, low consumption and broadband real-time RF spectral analyser. To overcome the electronic bottleneck encountered by electronic solutions, which limits the real time bandwidth to hundreds of MHz, we propose a new approach exploiting the quantum properties of the nitrogen-vacancy (NV) center in diamond. Here we describe a Quantum Diamond Signal Analyser (Q-DiSA) platform and characterize its performances. We successfully detect RF signals over a large tunable frequency range (25 GHz), a wide instantaneous bandwidth (up to 4 GHz), a MHz frequency resolution (down to 1 MHz), a ms temporal resolution and a large dynamic range (40 dB). 
\end{abstract}

\section*{Introduction}
The real time detection and the broadband spectral analysis of microwave (MW) signals is key for a broad range of technologies such as communication, medicine and navigation. Cognitive radio networks, electromagnetic compatibility (EMC) analysis, radars, wireless communications, etc. are some of the many applications that require a real time spectral detection over a broad frequency band (tens of GHz) with $100\%$ probability of intercept (POI).\newline
At present, the principal and most common solution for the real time spectral analysis of MW signals is the fast Fourier transform (FFT) electronic spectrum analyser \cite{RS}. The MW signal is digitalized by an analog-to-digital converter (ADC) and then processed by Fourier analysis. However, the sampling rate and the power consumption of the ADC limit the real time bandwidth to typically several hundreds of MHz. Increasing the analysis bandwidth can be achieved by analog solutions based on photonic approaches \cite{Zou2016}, where the MW signal is transposed in the optical domain and then processed to retrieve the spectral information. One promising technique is the spectral hole burning in ion-doped crystals at cryogenic temperature \cite{Louchet_Chauvet_2020}, with an instantaneous real time bandwidth of several tens of GHz, a frequency resolution of hundreds of kHz and a large dynamic range ($> 50$ dB) \cite{Merkel2016}. However, although cooling technique has made impressive progress with closed cryocoolers now available, there is a need for complementary solutions that are compact, can be operated at room temperature and require low power consumption, in particular for onboard components.\newline
An alternative technique is to use the quantum spin properties of electrons to directly detect MW signals with neither analog-to-digital conversions nor optical processing modules. The Nitrogen-Vacancy (NV) center in diamond is very attractive for this purpose since it is a solid-state atom-like system that can be utilized at room temperature. Its simplicity of operation makes it an appealing platform for building measurement devices, such as magnetometers \cite{Rondin_2014}, gyroscopes \cite{Ajoy2012}, thermometers \cite{Kucsko2013}, high-pressure sensors \cite{lesik2019}, electrometers \cite{Chen2017}, etc., that are compact, consist of simple optical components and that require low power consumption.\newline
Here we propose a Quantum Diamond Spectrum Analyser (Q-DiSA) for MW signals exploiting the properties of NV centers in diamond.
The proof of principle of NV-based MW spectral analysis has been demonstrated over a frequency range of some hundreds of MHz \cite{Chipaux2015}. In this work, we investigate the main physical features of NV centers considered as MW detectors, in order to extend the frequency range, the spectral resolution, the power detection threshold and the temporal resolution of the Q-DiSA technique. We demonstrate a practical architecture showing that NV centers are well suited for the real time spectral analysis over a frequency band of 25~GHz and we characterize its performances. 

\section*{The nitrogen-vacancy center}
The NV center is a color center in diamond \cite{DOHERTY2013} consisting of a Nitrogen atom and a Carbon vacancy in two adjacent positions of the diamond lattice (Figure 1a). The negatively charged state (NV$^-$,  now-on simply denoted as NV) with two unpaired electrons has an electron spin equal to 1. Due to the $C_{3V}$ symmetry of this defect, the N-to-V crystallographic direction forms an intrinsic spin quantization axis. According to the tetrahedral structure of the diamond lattice, four different N-to-V axis orientations exist in the crystal.\newline 
Due to its remarkable spin-dependent optical properties, the NV center is an enabling tool for numerous quantum sensing experiments \cite{Schirhagl2014}. NV energy levels are depicted in Figure 1b. Both ground state and excited state are spin triplets ($\ket{m_s=0,\pm1}$). The ground state zero field splitting (ZFS) between the $\ket{0}$ state and the two degenerate  $\ket{\pm1}$ states is equal to 2.87 GHz. At room temperature, these three spin states are equally populated due to thermal excitation.\newline
Under optical pumping, the main de-excitation process consists in spin-conserving radiative transitions characterized by a photoluminescence (PL) emission spectrum with a zero phonon line (ZPL) at $637$ nm and a broad electron-phonon band (up to $800$ nm). An additional non-radiative intersystem crossing (ISC) connects the $\ket{\pm1}$ excited state sublevels to the $\ket{0}$ ground state sublevel through a metastable singlet state \cite{Tetienne_2012} leading to an efficient polarization of the spin in the $\ket{0}$  state after a few optical pumping cycles, and a reduction of the PL rate when the $\ket{\pm1}$ sublevels are populated. \newline
Applying an external static magnetic field to the NV center leads to a Zeeman shift of the $\ket{0}\rightarrow\ket{+1}$  and $\ket{0}\rightarrow\ket{-1}$ radio frequency (RF) transitions. When the field is applied along the NV center axis, the two resonance frequencies are (Figure 1c):

\begin{equation}
\nu_\pm=|D\pm\gamma B_{NV} |
\end{equation}

\setlength{\parindent}{0pt} where $D$ is the zero field splitting equal to $2.87$ GHz, $B_{NV}$  is the magnetic field component parallel to the NV axis and $\gamma$, equal to $28$ GHz/T , is the NV center gyromagnetic ratio. In particular, for a magnetic field of $102$ mT, the $\ket{-1}$ level crosses the $\ket{0}$ level. \newline
These two magnetic resonances can be optically detected (ODMR) by sweeping the frequency of a RF signal while monitoring the PL level (Figure 1d). In this case, the NV center works as a magnetometer \cite{hong_2013}.\newline
The Q-DiSA technique consists in the reciprocal scheme. It relies on the spatial encoding of the NV center resonance frequencies by means of a controlled magnetic field gradient. The schematic is reported in Figure 1d. NV centers located at different positions into the diamond undergo a different Zeeman shift and thus resonate at different MW frequencies. The MW signal to be analyzed is sent on the diamond through an antenna while a camera collects the PL emitted by the NV centers. The decrease in the PL intensity on some pixels reveals the presence of spectral components of the MW signal that are resonant with the NV centers imaged by those pixels. The measurement is performed in a continuous wave (CW) regime, namely the laser pumps continuously the NV centers, allowing, at the same time, a continuous repolarization of the spin and an instantaneous spin read out, without any dead time for the signal detection, except the refreshing time of the camera. Therefore, the spectral components of the signal are measured, instantaneously, simultaneously and without any signal processing, by simply recording the image of the PL emitted by the diamond sample \cite{Chipaux2015}.
\begin{figure}
	\centering
	\includegraphics[width=0.9\linewidth]{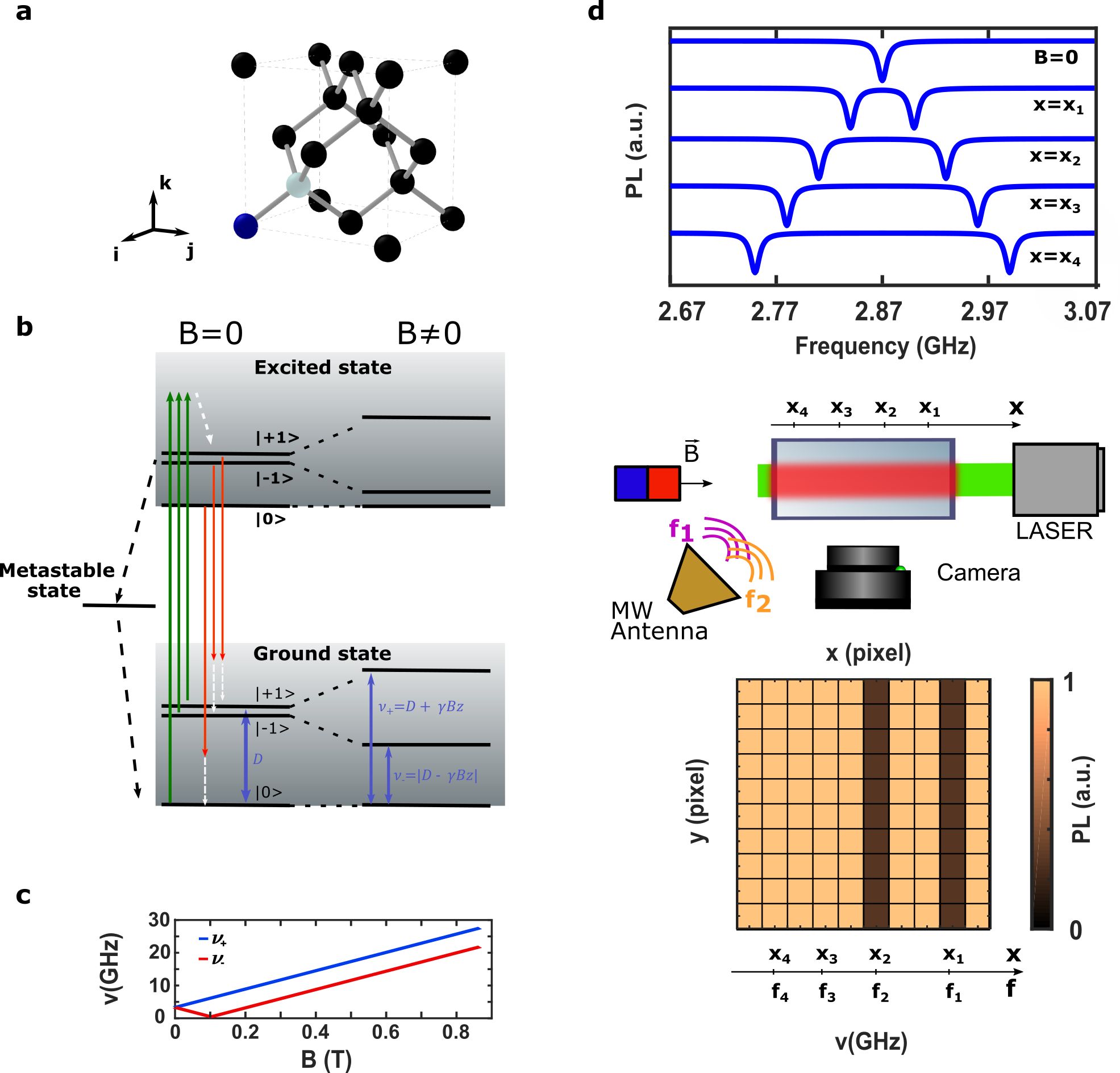}
	\caption{Nitrogen-Vacancy center schematic and its application to the spectral analysis. (a) NV center defect into the diamond lattice. (b) NV center energy levels. The system is optically excited by a non-resonant $532$ nm laser (green arrows). The main de-excitation process is a spin conserving radiative transition (red arrows). A non-radiative process (black dashed arrows) also occurs and allows for intersystem crossing (ISC). Both the excitation process and the radiative process can be phonon-assisted (white dashed arrows). (c)  NV center resonance frequencies induced by the Zeeman effect.  (d) General application of NV centers to the spectral analysis of MW signals. At the center, the schematic of the set-up. At the top, the ODMR spectra of NV centers placed at different distances from the magnet. At the bottom, the output of the camera when a MW signal is detected. A drop of photoluminescence is detected on the pixels that image the NV centers resonant with the applied signal.}
	\label{fig:fig1}
\end{figure}

\section*{Experimental set-up}
Using NV centers for microwave detection over a large frequency range requires both a strong static magnetic field and a magnetic field gradient.
\begin{figure}
\centering
\includegraphics[width=0.9\linewidth]{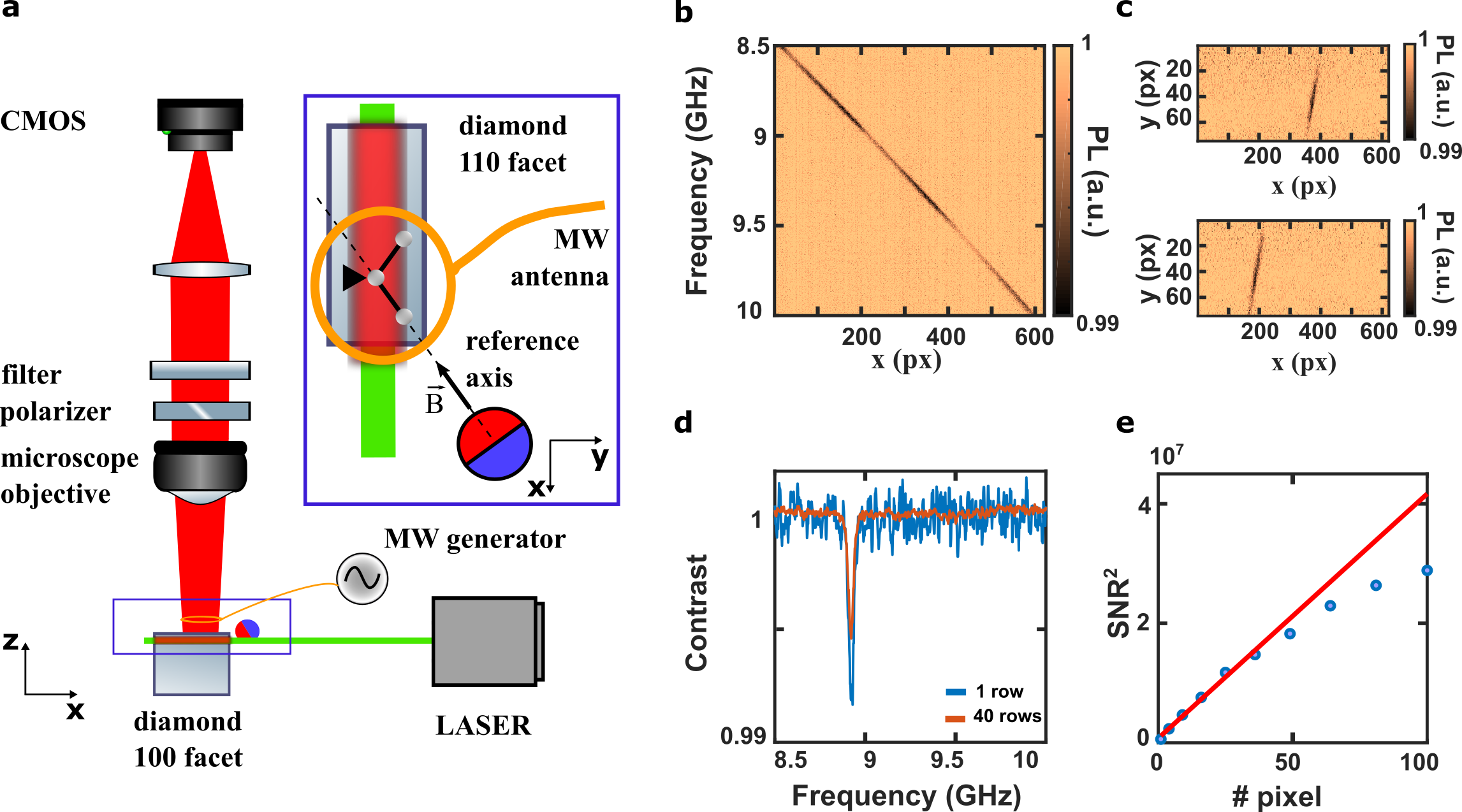}
\caption{Q-DiSA set-up and calibration procedure. (a)  Q-DiSA set-up. The inset shows a zoom of the 110 facet from which the PL is collected. The magnetic field B generated by the spherical magnet is aligned along one of the two NV center families laying on the 110 plane. (b) CW-ODMR of the $\ket{0}\rightarrow\ket{-1}$ transition for a magnet-diamond distance of approximately 2~mm. The resonance frequencies of the imaged NV centers are in the range [$8.5$~GHz; $10$~GHz]. 
(c) Normalized PL acquired by the camera in the presence of a monochromatic MW signal at 9.5~GHz (top) and 9~GHz (bottom). (d) Comparison between the spectrum obtained considering only one row of pixels, (y=45, blue curve) and a sum over 50~rows (y from 10 to 60, orange curve) showing a strong SNR improvement resulting from the summing procedure. 
(e) SNR of the PL integrated over a different number of non-resonant pixels. The signal is defined as the temporal mean of the PL; the noise is its standard deviation. The SNR increases linearly when the PL is integrated over a small number of pixels; the non linear behaviour of the SNR for integration areas bigger than 40~pixels is mainly due to the inhomogeneity of the laser beam.}
\label{fig:fig2}
\end{figure}
Neodymium magnets provide a compact and room temperature solution to fulfill those requirements. We used a magnetic sphere of 1.3~cm diameter (Supermagnet K-13-C) which provides a magnetic field in the range of 1~T.  By adjusting the magnet-diamond distance, the $\ket{0} \rightarrow \ket{-1}$ transition frequency can be tuned from 10~MHz to 21~GHz and the $\ket{0}\rightarrow\ket{+1}$ transition frequency from 2.87~GHz to 27~GHz (Figure~2a) (see Supplementary). In order to achieve such a high Zeeman shift and preserve the optical spin properties of the NV centers, the magnetic field has to be aligned carefully with the NV center axis \cite{Tetienne_2012}. The magnetization axis of the magnet is fixed in the horizontal plane and it is chosen as the reference axis of the system. The magnet is therefore mounted on a three-axis translation stage, with one horizontal axis parallel to the magnetization axis. The core element of the system is a commercially available $4.5 \times 4.5 \times 0.5$ mm $\{100\}$ single-crystal optical grade CVD diamond plate (ElementSix) doped with NV centers at a concentration of a few ppb and $\{110\}$ lateral facets which contain two NV axis in their plane. The diamond sample is mounted with a $\{110\}$ facet parallel to the horizontal plane, allowing us to align one of the two families along the magnetization axis of the magnet (see inset of Figure~2a). Therefore our system preserves the alignment of the magnetic field and the NV axis while adjusting the magnet-diamond distance (Figure~2a).\newline
The NV centers are optically excited using a cw laser at 532~nm wavelength. The laser is focused, with a full width at half maximum (FWHM) waist of  $38 \pm 3$~\textmu m, into the diamond plate through one of the four ${110}$ facets. The PL is collected, through the top $\{110\}$ facet, using an optical microscope objective (20~$\times$, 0.33 NA), spectrally filtered (FF01-$697/75$ nm band-pass filter, Semrock) and focused on a commercial CMOS camera (UI-$5240$CP-M, IDS) using a 75~mm lens. A polarizer, located after the objective, partially suppresses the PL emitted by the three NV centers families off-axis with the magnetic field in order to enhance the ODMR contrast of the on-axis family (Supplementary).\newline
The NV electronic spin transitions are excited by the homogenous RF magnetic field  produced by a $1$ mm-diameter loop antenna connected to a MW signal generator (SMA100B, R$\&$S).\newline
The optical system (beam waist and objective magnification) and the size of the antenna define the active area on the diamond ($530\times50$  \textmu m$^2$) and the spatial resolution ($0.66\times0.66$ \textmu m$^2$) (see inset of Figure~2a). We have chosen those parameters as the best compromise between the size of the PL image, giving the spectral bandwidth, and the spatial resolution (three times the diffraction limit) which defines the frequency resolution of the spectral analysis.\newline

\section*{Calibration procedure}
For a given magnet-diamond distance, the Q-DiSA calibration procedure consists in associating each pixel of the image to the corresponding NV resonance frequency. As illustrated on Figure~2b, the ODMR spectra show that, for a given y position, the spectral information is spatially coded along the x-axis through the magnetic field gradient (see inset of Figure~2a).\newline 
Instead of considering a single row of pixels, we can take advantage of the wide field imaging mode. Considering all the pixels that resonate at the same frequency, corresponding to the iso-B of the static magnetic field (Figure 2c), we create a one-to-one calibration map that links each frequency to a set of pixels. \newline
In wide field imaging systems the signal-to-noise ratio (SNR) is primarily affected by photon shot noise. We can therefore increase it (Figure 2d-2e) by summing over all the pixels which resonate at the same frequency.  The SNR then reads as \cite{Chipaux2014}:

\begin{equation}
\label{eq:SNR}
SNR=\sqrt{nVN_pR_0\zeta\Delta t}C
\end{equation}

\setlength{\parindent}{0pt}where $n$ is the NV center concentration of the diamond, $V$ is the volume imaged by the pixel, $N_p$ is the number of pixel resonating at the same frequency, $R_0$ is the NV center PL rate, $\zeta$ is the collection efficiency of the system, $\Delta t$ is the integration time of the measurement and $C$ is the ODMR contrast. Once the calibration is performed, the only signal processing needed to retrieve the spectral content of the RF signal of interest is a sum over a few pixels. This procedure is quick and independent of the frequency we want to detect.

\section*{Results}
Key parameters for the spectral analysis of MW signals are the frequency range, the bandwidth, the frequency resolution, the dynamic range and the temporal resolution. We discuss the performances of the Q-DiSA architecture with respect to these parameters. 
\subsection*{Frequency Range and Bandwidth}
The Q-DiSA system works as a tunable MW frequency detector in a typical range from 10~MHz to 25~GHz (Figure 3a). The central frequency is determined by the distance between the magnet and the diamond sample, while the bandwidth is determined by the magnetic field gradient. Due to the choice of the single-magnet configuration these two parameters are linked (Figure~3a). 
\begin{figure}
\centering
\includegraphics[width=0.9\linewidth]{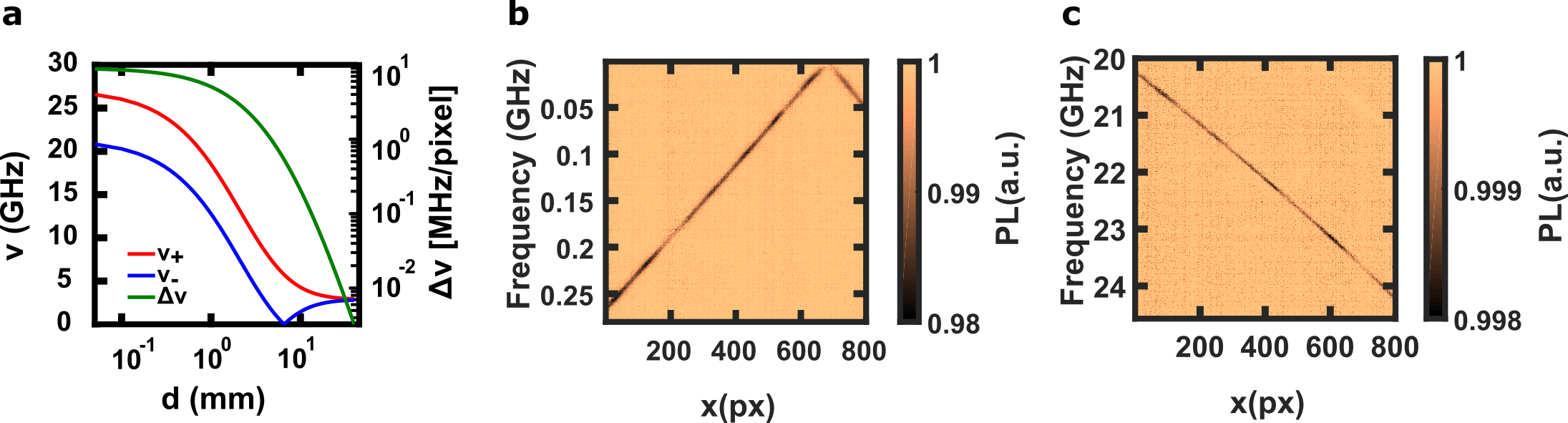}
\caption{Frequency range and bandwidth. (a) Ground state NV center resonance frequencies (blue and red) and magnetic field gradient (green) at different distances from the surface of the spherical magnet. The magnetic field gradient is expressed in MHz/pixel. (b) ODMR of the $\ket{0}\rightarrow\ket{-1}$ transition at the GSLAC. The field of view corresponds to a bandwidth between 10~MHz and 250~MHz. (c) ODMR of the $\ket{0}\rightarrow\ket{+1}$ transition around 22~GHz. The field of view corresponds to a bandwidth between 20~GHz and 24~GHz.}
\label{fig:fig3}
\end{figure}
The lower part of the spectrum is covered when the energy difference between the $\ket{0}$ and $\ket{-1}$  states becomes very small. In that case, even very weak couplings (such as parasitic strain in the diamond sample, hyperfine coupling, residual transverse magnetic field) can induce a mixing between the $\ket{0}$ and $\ket{-1}$ states. This mixing induces a ground state level anti-crossing (GSLAC) that blurs the correspondence between the frequency of the signal applied to the NV centers and the magnetic field gradient. This parasitic effect \cite{Ivady2021, Broadway2016} sets the lower detectable frequency of our sensor to typically  10~MHz (Figure~3b). The upper limit that we have reached is 27~GHz and it is related to the maximum amplitude of the static magnetic field produced by the permanent magnet (Figure 3c) (see Supplementary). These parameters could be extended using a stronger magnetic field that can be produced by cryogenic magnets \cite{Stepanov2015,Babunts2020}. \newline
The relation between the bandwidth and the strength of the magnetic field (Figure 3a) is given on Figure 3b-3c showing, for the same diamond area, a total bandwidth of  300~MHz  at a central frequency of 150~MHz (Figure~3b) and a bandwidth of 4~GHz at a central frequency of 22~GHz (Figure~3c). Those values can be adapted by changing both the imaging system and the magnetic field architecture in order to tune the amplitude and the gradient of the magnetic field as two independent parameters. 

\subsection*{Frequency resolution}
The frequency resolution of the Q-DiSA architecture is related to the resonance linewidth of the NV centers ensemble, which is affected by three main contributions: the intrinsic linewidth, the power broadening induced by both the optical and the MW excitation, and the inhomogeneous broadening induced by the magnetic field gradient \cite{Bauch2018}.\newline
In our optical grade diamond sample, the intrinsic linewidth, which can be attributed to the spin decoherence induced by the interaction between the NV centers and the $^{13}C$ nuclear spin bath \cite{Maze2008}, is of the order of 500 kHz. \newline
The other two contributions to the line broadening are shown in Figure 4a and Figure 4b, where the ODMR spectra for the same frequency set-point but under different magnetic field configurations (10~mT and 20~kHz/\textmu m (Figure 4a) – 195~mT and 1.5~MHz/\textmu m (Figure 4b)) are reported. In the presence of a low magnetic field gradient and for low  power MW excitation, the NV center hyperfine structure is well resolved (Figure 4a). It consists of three lorentzian peaks separated by 2.14~MHz \cite{Doherty2012} resulting from the hyperfine interaction between the NV center electron spin and the nitrogen atom nuclear spin ($^{14}N$, S=1). The linewidth of each peak is of the order of 1~MHz, which sets the best frequency resolution we are able to achieve with our system. The deviation from the intrinsic linewidth is attributed to the optical power broadening induced by the laser (Supplementary). Increasing the MW power, the ODMR line shows the typical power broadening effect and the linewidth is proportional to the amplitude of the MW field, as illustrated in Figure~4c. At low MW power and higher magnetic field gradient (Figure~4b), the hyperfine structure cannot be resolved anymore since the magnetic field gradient induces a distribution of the NV centers resonance frequencies at the single pixel scale thus broadening the ODMR line (Supplementaries).
Therefore, as clearly shown in Figure 4c, the frequency resolution of our system, at low power, is limited by the inhomogeneity of the static magnetic field and by some possible imperfections of the set-up that occur during the measurement time, such as mechanical vibrations and temperature fluctuations of the magnet \cite{Broadway2018}. Consequently, due to the frequency-dependent magnetic field gradient implemented in this single-magnet architecture, the frequency resolution is higher at low frequencies (FWHM of 1~MHz at 2.6~GHz), close to the ultimate resolution set by the NV center intrinsic linewidth, than at high frequencies (FWHM of 50~MHz at 21~GHz). 
\begin{figure}
\centering
\includegraphics[width=0.9\linewidth]{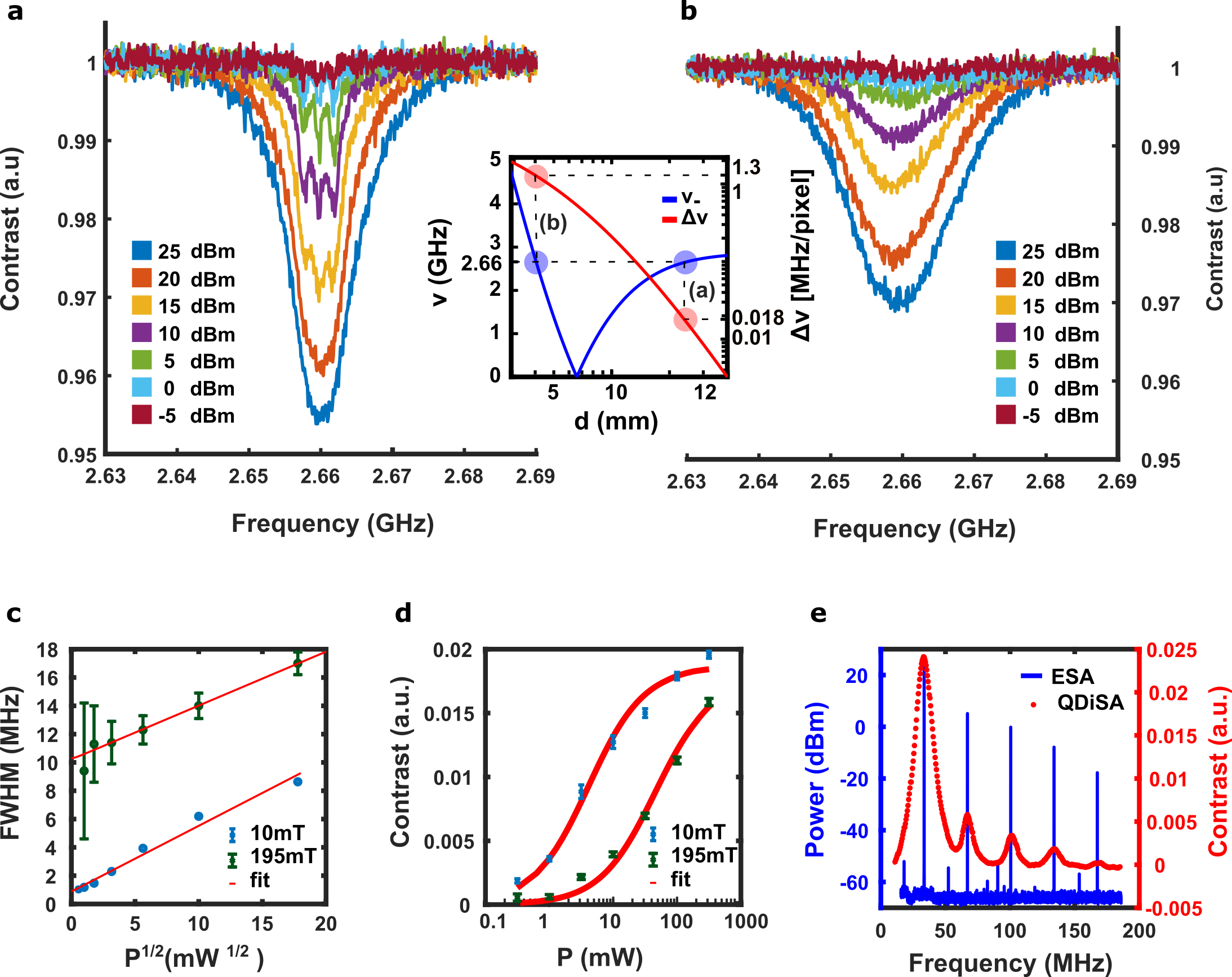}
\caption{Comparison of the ODMR spectrum in the range [2.63~GHz, 2.69~GHz] for two different magnetic fields and magnetic field  gradients (10~mT and 20~kHz/\textmu m  (a) – 195~mT and 1.5~MHz/\textmu m (b)). The curves are fitted using a sum of three identical Lorentzians shifted by 2.14~MHz accounting for the hyperfine structure of the resonance line to retrieve the FWHM (c) and the contrast (d) from the spectra. 
(e) Spectrum of the non linearity of the MW generator when it is set to a central frequency of 38~MHz and a nominal power of 25~dBm. }
\label{fig:fig4}
\end{figure}

\subsection*{Amplitude and dynamic range}
As shown in Figure~4d, the contrast increases with respect to the MW power, exhibiting a linear behaviour at low power regime and a saturation for stronger MW field (Supplementary). This effect can be used to quantitatively measure the amplitude of the magnetic field associated to the MW signal, as reported in \cite{Dreau2011,Shao2016}. The power detection threshold (i.e. the smallest contrast that can be detected) is given by the SNR (Eq.\eqref{eq:SNR}) of the measurement.  The differences in the ODMR contrast between the spectra obtained for a magnetic field equal to 10~mT (Figure~4a) and 195~mT (Figure~4b) are caused by the gradient of the static magnetic field, for the same reasons previously explained regarding the ODMR linewidth. Note that the MW power level used in Figure~4 is the nominal power level at the input port of the MW transmission line and not the MW power applied to the NV centers, which depends on the frequency-dependent transmission efficiency of the antenna. Therefore for the same MW nominal power, the contrast is higher at low frequencies than at high frequencies. \newline
As an example, the Q-DiSA is applied to characterize the non-linearity of a MW generator (SMA100A, R$\&$S). The results are reported in Figure~4e and compared to the spectrum measured using a commercial electronic spectrum analyzer. Considering the noise floor of the measurement, the Q-DiSA can detect a signal over a MW power range from 23~dBm to -17~dBm which opens the way to devices with a dynamics up to 40~dB.

\subsection*{Time resolution}
A key advantage of the CW wide-field imaging mode is that all the frequencies within the frequency range are simultaneously detectable without any dead time except the one of the camera. This spatial multiplex advantage makes possible the real time frequency detection of complex microwave signals with $100\%$ POI. To illustrate this point, a time varying multi-frequencies microwave field is generated by combining the signals of two microwave generators and by sweeping their frequencies over the entire bandwidth of the image (Figure~5a), for two different configurations, around 7~GHz (Figure~5b) and around 23~GHz (Figure~5c). The results indicate that simultaneous detection is possible for a number of frequency channels determined by the ratio between the bandwidth and the frequency resolution. \newline
The temporal resolution (\textDelta t) is the smallest time required to detect the variation of the PL signal induced by a resonant MW frequency component. Using Eq.\eqref{eq:SNR}, we evaluate the temporal resolution by decreasing the camera exposure time in order to reach values of the SNR close to 1, not considering the refreshing time of the camera (Figure 5d-5e-5f). Using a nominal MW power of 25 dBm, we measured a temporal resolution of 2~ms at 1.8~GHz (C=6$\%$, SNR=$5$), 20~ms at 9~GHz (C=$1\%$, SNR=4) and 600~ms at 23~GHz (C=$0.1\%$, SNR=$1$). This strong variation of the temporal resolution with frequency is a consequence of the contrast frequency dependence. It is directly related to the Q-DiSA architecture and not to the intrinsic NV center photodynamics, whose characteristic times are of the order of one microsecond \cite{Tetienne_2012,robledo2011}.
\begin{figure}
\centering
\includegraphics[width=0.9\linewidth]{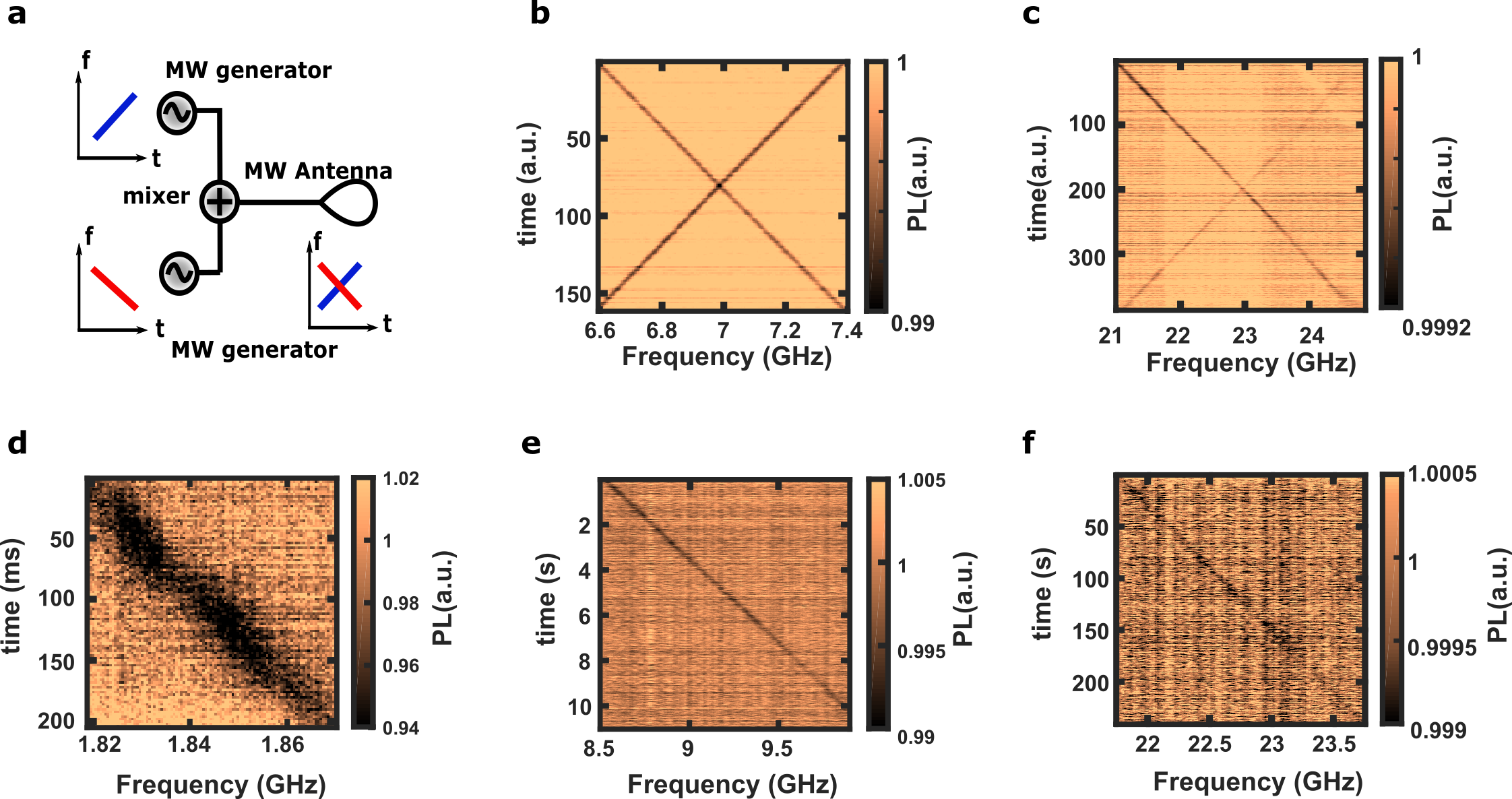}
\caption{Simultaneous detection and temporal resolution. (a) Schematic of the multi-frequencies signals detection. (b-c) Multi-frequencies signals detection at 7~GHz (b) and 23~GHz (c). (d-f) Spectrogram for three different frequency ranges: 1.8~GHz (d), 9~GHz (e), 23~GHz (f)}
\label{fig:fig5}
\end{figure}

\subsection*{Frequency ambiguity}
The Q-DiSA allows analysing the frequency components of a MW signal over a wide range. However, it may present some frequency ambiguities, namely the presence of more than one resonance frequency associated to the same pixel. Two factors can induces such ambiguity: the presence of additional ODMR lines due to the three non-aligned NV families and the two intrinsic resonance frequencies of the NV center ($\ket{0}\rightarrow\ket{-1}$ and the $\ket{0}\rightarrow\ket{+1}$). The first source of ambiguity can be eliminated using a diamond with NV centers preferentially orientated along one of the four diamond crystallographic axis \cite{Miyazaki2014}. The second source is intrinsic to the single NV center but can be suppressed by applying a MW filter so that the signal is resonant with only one of the two NV center transitions. This procedure restricts the available bandwidth to a maximum of 5.74~GHz (twice the zero-field splitting), corresponding to the frequency shift between the $\ket{0}\rightarrow\ket{-1}$ and the $\ket{0}\rightarrow\ket{+1}$ transition for magnetic fields stronger than 0.1~T (Figure 1c). In this configuration the frequency range can be chosen according to the best performances of the MW chain (for example corresponding to baseband communication), and any other frequency signal can be brought inside this frequency range using heterodyning technique. Such mixing will then lead to a spectral analysis over a frequency range larger than 25~GHz, only limited by the heterodyning chain.   

\section*{Conclusion}
We demonstrated that an ensemble of NV centers in a single-crystal diamond can be used to detect a complex microwave signal and resolve its spectral components. The analysis bandwidth can be tuned from a few MHz up to 25~GHz. The Q-DiSA experimental platform is compact and is operated at room temperature with low power consumption. The performances achieved in this work, corresponding to MHz frequency resolution and ms temporal resolution, can be improved working both on the experimental architecture (e.g. using an improved optical collection, a more flexible magnetic field architecture, different pumping parameters, etc.) and on the diamond sample (e.g. using a higher concentration of NV centers \cite{Edmonds2021} or a crystal with NV centers having a preferential orientation).
Moreover, the large frequency range of the Q-DiSA can be further extended using heterodyne techniques, opening the way to the real time spectral analysis over a frequency range of several tens of GHz.

\section*{Methods}
\subsection*{NV center optical excitation}
The 532 nm laser is set to 300 mW. The effect of the laser power to the ODMR linewidth are discussed in the supplementary materials.
According to \cite{Wee2007}, the NV center saturation intensity is: I\textsubscript{Sat}=1-3 GW/m$^2$. Considering a laser beam waist $w_{0} \approx40$~ \textmu  m (as it is in our experimental architecture), the saturation power is achieved at $P_{Sat} = \frac{\pi w_0^2}{2}I_{Sat}\approx$ 2~W. Therefore, using a laser power of 300 mW, the saturation parameter is $s=P/P_{Sat}=0.15$ (see supplementary materials).\newline

\subsection*{The loop antenna}
The loop antenna employed to bring the RF field in proximity of the NV centers is realized short-circuiting a co-axial cable with a 50~\textmu m diameter copper wire which is soldered to the coaxial cable.
The microwave field generated inside the loop is almost homogenous and linearly polarized along the loop axis. Placing the antenna at the top of the diamond [110] facet, the RF field is perpendicular to the NV center family aligned to the reference axis, thus maximizing the coupling between the RF field and the electron spin.

\subsection*{The ODMR spectrum acquisition procedure}
In this section we detail the acquisition procedure used to acquire the ODMR spectra reported in Figure 2b-2c, Figure 3b-3c, Figure 4a-4b.
It consists in sweeping the frequency of the RF field while the camera captures, for each frequency, the PL emitted by the NV centers. During the acquisition the laser and the RF field are always on.
In order to speed up the acquisition procedure, the area of interest (AOI) of the camera is limited to the pixels that image NV centers pumped by the green laser.\newline
The RF frequency $f$ is scanned in a range $\left[f_{min},f_{max}\right] $ with a frequency resolution of $\Delta f$, which results in $N_{seq}$ = ($f_{max}$ - $f_{min})/\Delta f$ frequency steps. In order to increase the SNR of the measurement, the frequency ramp is repeated $N_{cycle}$ times; for each frequency step, the signal acquired by the camera is added to the signal acquired during the previous ramps at the same frequency.
The number of cycles over which the RF frequency is swept is chosen in order to obtain a satisfactory SNR. The exposure time of the camera is chosen in order to maximize the number of counts without saturating the detector.\newline
During the acquisition, the data are stored in a three dimensional matrix of size [w; h; $N_{seq}$], where w and h are respectively the width and the height of the AOI. 
In order to retrieve the contrast of the measurement the data are normalized dividing the data matrix (D) for the mean value of the PL in the first m and the last m steps of the frequency ramp, checking that all the NV centers in the AOI are out of resonance for those frequency values. The normalized data matrix (D\textsuperscript{Norm}) is thus given by:

\begin{equation*}
D^{Norm}_{xyf}=\frac{2m\cdot D_{xyf}}{\sum_{f=f_{min}}^{f=f_{min}+m \Delta f}D_{xyf}+\sum_{f=f_{max}-m \Delta f}^{f=f_{max}}D_{xyf}}
\end{equation*}

and it is directly related to the ODMR contrast by the formula:

\begin{equation*}
C_{xyf}=1-D^{Norm}_{xyf}.
\end{equation*}

\subsection*{The calibration procedure}
The calibration procedure consists in defining, for each frequency in the frequency range, a mask of the same size of the camera AOI, which is zero everywhere except for the pixels which resonate at the frequency associated to the mask. Once the set of calibration masks has been built, the spectrum is defined as:

\begin{equation*}
	S(\nu)=\sum_{x,y}I(x,y)\times M(x,y,\nu)
\end{equation*}

where $I(x,y)$ is the image acquired by the camera when the signal we want to analyse is sent in proximity of the NV center and $M(x,y,\nu)$ is the mask associated to the frequency $\nu$.

\subsection*{Fitting procedure}
The fit curve used to fit Figure 4a-4b and thus to retrieve the ODMR contrast and ODMR linewidth plotted in figure 4c-4d is given by:
\begin{equation*}
	f(\nu)=1-\left(\frac{a^2}{a^2+(\nu-b)^2}+\frac{a^2}{a^2+(\nu-b+\nu_{hyp})^2}+\frac{a^2}{a^2+(\nu-b-\nu_{hyp})^2}\right)c
\end{equation*}
where $\nu$ is the frequency expressed in MHz, $\nu_{hyp}$~=~2.14~MHz is the frequency difference between two successive peaks of the NV center hyperfine structure, a is a parameter related to the ODMR linewidth, b is a parameter related to the NV center resonance frequency and c is a parameter related to the contrast. 
The FWHM is then defined as $FWHM=2a$ and the contrast as $C_{ODMR}=c$. The error bars of figure 4c-4d are due to the error on the fit parameters considering a confidence interval of $95\%$. 

\section*{Acknowledgments}
This project has received funding from the European Union’s Horizon 2020 research and innovation programme under grant agreement No. 820394 (ASTERIQS), the Marie Skłodowska-Curie grant agreement No. 765267 (QuSCo) and the QUANTERA grant agreement ANR-18-QUAN-0008-02265 (MICROSENS). We acknowledge the support of the DGA/AID under grant agreement ANR-17-ASTR-0020 (ASPEN).

\bibliographystyle{unsrt}
\bibliography{paper}

\end{document}